# *Datasets of Great Britain Primary Substations Integrated with Household Heating Information*

**Authors**

*Yihong Zhou[1]\*, Chaimaa Essayeh[2], and Thomas Morstyn[3]*

**Affiliations**

*1. School of Engineering, University of Edinburgh, Edinburgh, U.K., EH9 3FB*

*2. Department of Engineering, Nottingham Trent University, Nottingham, U.K., NG1 4FQ*

*3. Department of Engineering Science, University of Oxford, Oxford, U.K., OX1 3PJ*

*\*Correspondence: [yihong.zhou@ed.ac.uk](mailto:yihong.zhou@ed.ac.uk)*

**Abstract**

The growing demand for electrified heating, electrified transportation, and power-intensive data centres challenge distribution networks. If electrification projects are carried out without considering electrical distribution infrastructure, there could be unexpected blackouts and financial losses. Datasets containing real-world distribution network information are required to address this. On the other hand, social data, such as household heating composition, are closely coupled with people's lives. Studying the coupling between the energy system and society is important in promoting social welfare. To fill these gaps, this paper introduces two datasets. The first is the main dataset for the distribution networks in Great Britain (GB), collecting information on firm capacity, peak demands, locations, and parent transmission nodes (the Grid Supply Point, namely GSP) for all primary substations (PSs). PSs are a crucial part of the UK distribution network and are at the lowest voltage level (11 kV) with publicly available data for most UK Distribution Network Operators (DNOs). Substation firm capacity and peak demand facilitate an understanding of the remaining room of the existing network. The parent GSP information helps link the dataset of distribution networks to datasets of transmission networks. These datasets are collected, processed, and merged from various files published by the six DNOs in GB. There were inconsistencies among the PS names across files even for the same DNO. A Python script and manual validation are performed to carefully process and merge the corresponding PS information. The second dataset extends the main network dataset, linking each PS to information about the number of households that use different types of central heating recorded in census data (Census in year 2021 for England and Wales, and Census 2011 for Scotland as the most recent Scotland Census 2022 data has not yet been fully released). The derivation of the second dataset is based on locations of PSs collected in the main dataset with appropriate assumptions. The derivation process may also be replicated to integrate other social datasets.

# SPECIFICATIONS TABLE

| | |
|---|---|
| **Subject** | Electrical and Electronic Engineering |
| **Specific subject area** | The datasets aggregate information about distribution networks in Great Britain and integrate the network information with household heating information. |
| **Type of data** | Table |
| **Data collection** | All energy system data is collected from the publicly available websites maintained by the six Distribution Network Operators in Great Britain (England, Scotland, and Wales). All housing data and geographical data are sourced from the UK Census and government websites. |
| **Data accessibility** | Repository name: Zenodo<br>Data identification number: DOI: 10.5281/zenodo.10866260<br>Direct URL to data: https://doi.org/10.5281/zenodo.10866260 |
| **Related research article** | Zhou, Yihong, Chaimaa Essayeh, Sarah Darby, and Thomas Morstyn. "Evaluating the social benefits and network costs of heat pumps as an energy crisis intervention." iScience 27, no. 2 (2024), DOI: https://doi.org/10.1016/j.isci.2024.108854. |

# VALUE OF THE DATA

- The data contain information on primary substations in Great Britain, reflecting the characteristics of the UK distribution networks. The information includes location, demand, capacity, and the parent grid supply points. These data are beneficial for power grid evaluations considering the headroom of existing network infrastructure.
- The data contain the geographic location of each primary substation, which is then linked to the household heating data sourced from the UK census. Researchers can leverage these cross-disciplinary data to explore the intersection of energy management, social dynamics, and infrastructure planning, which facilitates understanding of the complex interactions involved.
- The data are for the entire Great Britain, which can facilitate nationwide research and policy-making projects.
- The introduced datasets can be extended to integrate other datasets, including the transmission network datasets based on the provided parent transmission network node information, as well as other social datasets using the provided location information and a process described in the Methods section.

# BACKGROUND

The growing demand for electrified heating, electrified transportation, and power-intensive data centres place a burden on distribution networks [1]. Electrification projects need to consider network infrastructure to avoid unexpected blackouts and financial losses. The existing dataset [2] for Great Britain (GB) did not provide information on network capacity and demand, which is insufficient for evaluating the feasibility of new load and generation connections. Furthermore, there is a trend towards system design which holistically considers new energy generation/storage/flexibility technologies, energy network infrastructure, and societal objectives, which requires more comprehensive datasets [3]. To fill the gaps, this paper releases two datasets. The first provides information on Primary Substations (PSs) of GB. PSs in the UK typically step down the voltage from 33 kV to 11 kV, which are a crucial part of the UK distribution networks and are at the lowest voltage level with publicly available data for most UK Distribution Network Operators (DNOs). This paper also provides an extended dataset that links household heating data to each PS, which helps illuminate the coupling between the energy system and society. The datasets supported the analysis in [1], and the information on two DNOs (SPEN and ENW) is updated to a more recent version.

# DATA DESCRIPTION

This article publishes two processed datasets at Zenodo [4]. To promote replication, the associated code to process the data as well as the raw data are released in [4] as well. Here, we first describe the key outputs, namely the two processed datasets. The first "GB_PS_data.csv" is the main network dataset, which provides the information on PSs extracted and processed from all the six DNOs in GB. The second "GB _PS_data_extended.csv" is denoted as the extended dataset, which is derived by estimation that links the census datasets to each PS in the main dataset. Both datasets are tables with 4,205 rows stored in CSV-format files. Each row represents a PS, with columns representing different aspects of information, which are described below.

## The Main Dataset

The main dataset "GB_PS_data.csv" contains 7 columns:

1. PS Name: the name of the PS recorded by the corresponding DNO.
2. Firm Capacity (MVA): According to the definition provided by Northern Powergrid (NPG, one of the UK DNOs) [5], the firm capacity of a substation is the capacity that is immediately available after the occurrence of a first circuit outage. Therefore, the firm capacity can well measure whether a PS can supply electricity over the peak load time reliably. The unit of the substation firm capacity is Mega Volt-ampere (MVA).
3. Demand (MVA): the half-hourly peak demand in MVA of the PS, net the contribution from the connected local generation.
4. PF: the Power Factor (PF), which is the ratio between the active power and the apparent power over the peak load time. As explained in Section Dataset Standardisation, only PSs of ENW and SPEN (two of the UK DNOs) have information to infer the PFs in our released main dataset. However, 95% of PSs have PFs greater than 0.95, so it is reasonable to assume an average PF equal to 0.989 for the remaining PSs.

5. Geo(Long,Lat): the geographical location of the PS represented by longitude and latitude in decimal degrees.
6. GSP: the parent Grid Supply Point (GSP, the connection node to the transmission grid) of the PS.
7. DNO: the DNO that manages this PS. There are 6 DNOs covering GB, including: Electricity Northwest (ENW), Northern Powergrid (NPG), SP Energy Network (SPEN), Scottish and Southern Electricity Networks (SSEN), UK Power Networks (UKPN), and National Grid Electricity Distribution (NG, formerly known as Western Power Distribution, namely WPD).

By using the processed "GB_PS_data.csv", we visualise the distribution of the GB PSs in Figure 1, with colours representing the ratio (%) of the peak demand to the substation firm capacity.

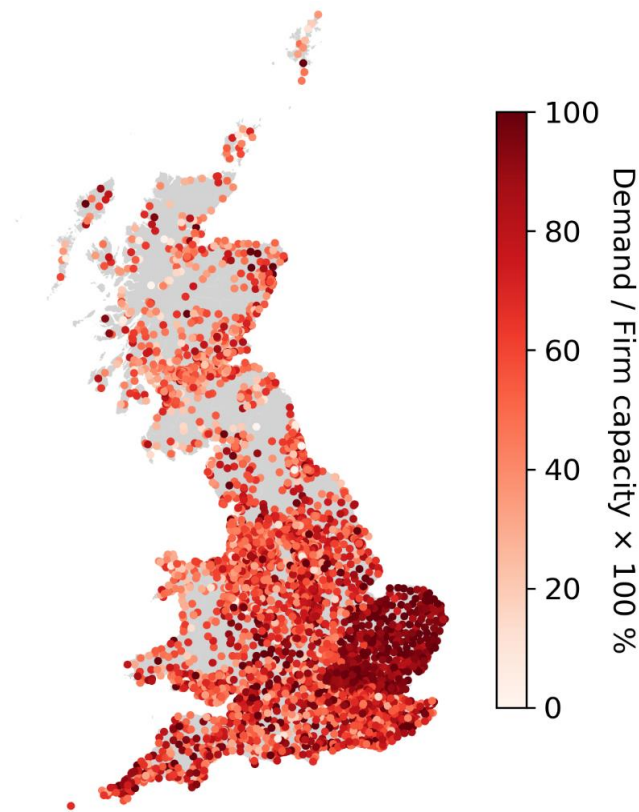

*Figure 1 Primary substations (PSs) in Great Britain (GB). The colour represents the ratio (%) of the peak demand to the substation firm capacity.*

### The Extended Dataset

The extended dataset "GB _PS_data_extended.csv" additionally contains the following attributes for each PS:

1. RegID: The digital code representing the sub-national region where the PS is located. As explained in Section Derivation of the Extended Dataset, the region is used for estimating the number of connected households for each PS. The region is the local authority (LA) for PSs in Scotland, the government region for PSs in England, or Wales as a whole.
2. RegName: the corresponding official name for the RegID.
3. Household no: The number of households connected to this PS, estimated by the method described in Section Derivation of the Extended Dataset.
4. NearestZone0. The digital code of the nearest small geographical zone of this PS, which is used to link the census household heating data as explained in Section Residential Heating

Composition. The geographical zone corresponds to the Middle Layer Super Output Area (MSOA) for England and Wales, and the Intermediate Zone (IZ) for Scotland, as explained in Section Residential Heating Composition.

5. NearestZone1. The digital code of the second-nearest geographical zone of the PS. The purpose of this feature is the same as "NearestZone0".

The following columns collect the household heating information, more specifically, the estimated number of households using different sources as the main central heating method, connected to each PS. For Scotland data, the total number of households summed over each heating option is equal to the provided total number of households; for England and Wales data, we calculate the total number of households summed over all heating options for each MSOA and compare it to another specific dataset for the total number of households for each MSOA, and observe a small deviation of 0.85% measured by the mean absolute percentage error (MAPE). Thus, we consider that households are not double-counted for different types of heating options.

6. No central heating: The estimated number of households without central heating.
7. Gas central heating: The estimated number of households that use natural gas as the main heating source. These homes are typically connected to the mains gas grid in the UK.
8. Electric (including storage heaters) central heating: The estimated number of households using electric heating, including (resistive) electric direct heating, (resistive) electric storage heating, and heat pumps. Note that, based on the most recent statistics [6], as of 2022 winter, only 1% of UK households were using heat pumps as the primary heating method.
9. Oil central heating: the estimated number of households using oil (mainly Kerosene) central heating.
10. Solid fuel (for example wood, coal) central heating: the estimated number of households using solid fuel central heating.
11. Other central heating: this records the estimated number of households not using any types of central heating listed here.
12. Two or more types of central heating: the estimated number of households using two or more types of central heating with types not specified.
13. District or communal heat networks only: the estimated number of households that use district or communal heat networks for heating.
14. Renewable energy only: the estimated number of households using renewable energy for heating. Examples include solar thermals or heat pumps, based on the explanatory notes on the England census questionnaire.
15. Tank or bottled gas only: the estimated number of households using tank or bottled gas for heating. One typical example of this type of heating source is the Liquefied Petroleum Gas (LPG).
16. Two or more types of central heating (including renewable energy): the estimated number of households using two or more types of central heating including renewable energy heating.
17. Two or more types of central heating (not including renewable energy): the estimated number of households using two or more types of central heating not including renewable energy heating.
18. Wood only: the number of households using wood for heating.

There are household heating options exclusively considered in Scotland or England/Wales. The heating option exclusively for Scotland is "Two or more types of central heating"; the options exclusively for England and Wales are: "District or communal heat networks only", "Renewable energy only", "Tank or bottled gas only", "Two or more types of central heating (including renewable energy)", "Two or more types of central heating (not including renewable energy)", and "Wood only". Therefore, missing values exist when a heating option is not modelled in the corresponding region of a PS.

## Codes and the Raw Data

Except the two processed datasets ("GB_PS_data.csv" and "GB_PS_data_extended.csv"), all folders and files in [4] correspond to the codes and the raw data. The structure tree of the whole repository is plotted in Figure 2. "GB_PS_data.csv" is the described main network dataset, and "GB_PS_data_extend.csv" is the extended dataset. "LICENSE" is the GPL-3.0 license file of the whole repository that allows free use, modification, and distribution. "link2social_data.ipynb" is the code script that integrates the main network dataset with the household heating data. "PS_data.ipynb" is the code script that processes and creates the main network dataset. "plot_map.ipynb" is the code file that plots Figure 1. "README.md" is a Markdown-based description of the whole repository. "requirements.txt" specifies the Python packages required to implement the codes. "GB_PS_utilisation_map.png" is the image displayed as Figure 1.

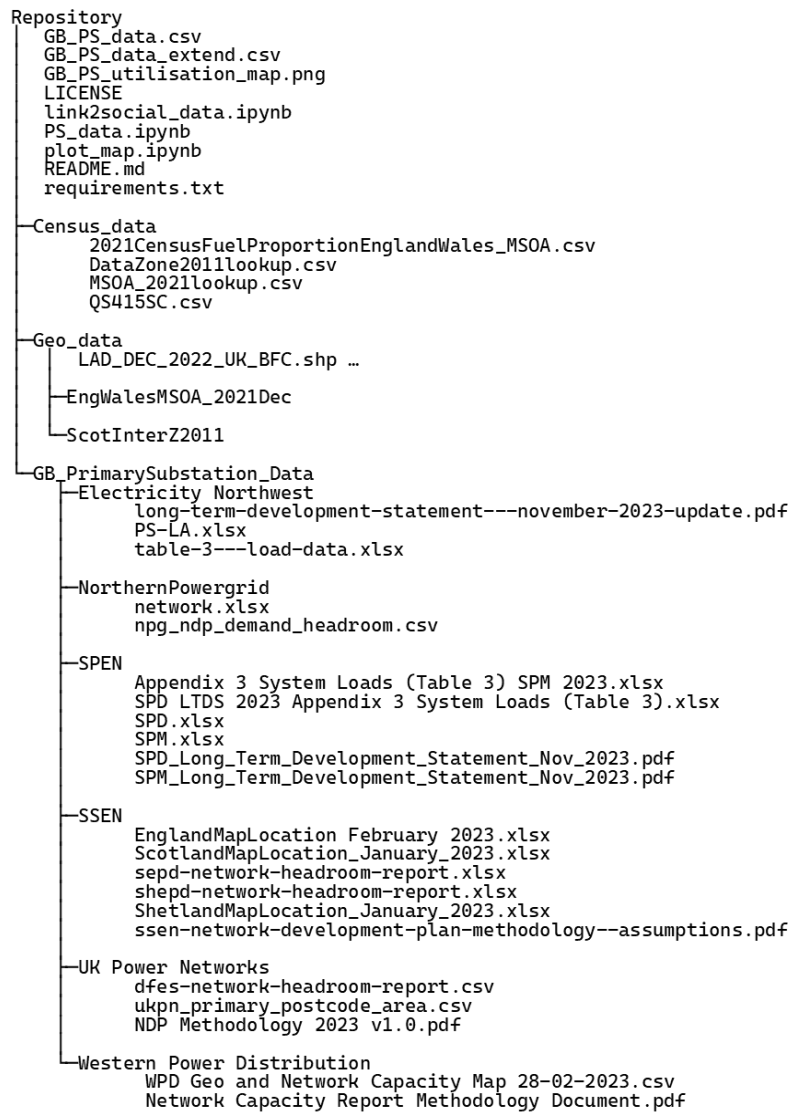

Figure 2. The tree structure of the released repository containing the two key processed datasets, the codes for data processing, and the raw data.

The folder "Census_data" includes all the household heating information collected by the UK census. "2021CensusFuelProportionEnglandWales_MSOA.csv" is the household heating data for England and Wales, while "QS415SC.csv" is that for Scotland. "DataZone2011lookup.csv" is the look-up table that links the different geographic levels in Scotland Census, while "MSOA_2021lookup.csv" is the look-up table for England and Wales.

The folder "Geo_data" includes all the geographical information of UK regions used to integrate the household information, as described in Section Derivation of the Extended Dataset:

- Files inside of "Geo_data" such as "LAD_DEC_2022_UK_BFC.shp" record the geographic boundaries of the LAs in Scotland, government regions of England, and the whole Wales.
- Two subfolders, "EngWalesMSOA_2021Dec" and "ScotInterZ2011", store locations of the population centroids of MSOAs in England and Wales, and IZs in Scotland.

The folder "GB_PrimarySubstation_Data" contains the source network data collected from the six UK DNOs listed in Table 1 (will be described in Section Collection of the Main Dataset). The raw data is stored in separate folders for each UK DNO. Here we map those files to items listed in Table 1:

- For the folder "Electricity Northwest", "PS-LA.xlsx" refers to the "Network Headroom Report data", while the other ".xlsx" file refer to the long-term development statement (LTDS) data in Table 1. "long-term-development-statement---november-2023-update.pdf" is the LTDS explanatory document.
- For "NorthernPowergrid", "network.xlsx" refers to the "Northern Powergrid Capacity Heatmap" data, and "npg_ndp_demand_headroom.csv" is the "NPG NDP Demand Headroom" data.
- For "SPEN", "Appendix 3 System Loads (Table 3) SPM 2023.xlsx" and "SPD LTDS 2023 Appendix 3 System Loads (Table 3).xlsx" are the "SPEN Long-term Development Statement" data. The remaining data files (.xlsx) are the manually extracted location information as discussed in Section Dataset Standardisation.
"SPD_Long_Term_Development_Statement_Nov_2023.pdf" and "SPM_Long_Term_Development_Statement_Nov_2023.pdf" are the LTDS explanatory documents.
- For "SSEN", "sepd-network-headroom-report.xlsx" and "shepd-network-headroom-report.xlsx" are the "Demand Headroom Report" data. "ssen-network-development-plan-methodology--assumptions.pdf" contains official explanation of terms and the data collection process. Other ".xlsx" files are the "Availability Map" data.
- For "UK Power Networks", "ukpn_primary_postcode_area.csv" is the "UK Power Networks PS area map" data, while the other ".csv" file is the "Distributed Future Energy Scenarios Network Scenario Headroom Report" data. "NDP Methodology 2023 v1.0.pdf" is the methodology document for the headroom report.
- For "Western Power Distribution", the ".csv" file is the "Network Capacity Map" data from the "NG" DNO in Table 1. "Network Capacity Report Methodology Document.pdf" is the methodology document for the network headroom report.

Please make proper reference to the primary data sources when the raw DNO data contained in our released data files in [4] are further utilised.

# EXPERIMENTAL DESIGN, MATERIALS AND METHODS

This section details the process for collecting and processing the main dataset and deriving the extended dataset.

## Collection of the Main Dataset

We summarise data from the six DNOs covering GB, including: ENW, NPG, SPEN, SSEN, UKPN, and NG. The sources of all the information are listed in Table 1. SPD and SPM are the two license areas of the distribution networks managed by SPEN. For each type of information, the corresponding unit (if applicable) is listed in parentheses. Here, "X, Y" represents the location recorded in the projected coordinate system for the UK (EPSG: 27700); "LongLat" represents the commonly used longitude/latitude system (EPSG:4326) in decimal degrees. All the source data used here reflect the network conditions in year 2022/2023.

## Dataset Standardisation

From Table 1, we can see that the same type of information in different DNOs have different formats. Standardisation is needed to facilitate the construction of the concatenated dataset for the whole GB. We consider the following standardised attributes:

1. The peak demand at a half-hourly resolution in MVA, net of the local generation contributions. The following DNOs' information needs to be processed:

   The peak demand recorded in SSEN is expressed in MW, which represents the active power after deducting the local generation. This attribute needs to be converted to MVA by using the PF over the peak load time. As listed in Table 1, only the ENW and SPEN datasets contain the PF information for each PS. Note that the DNO source data does not clearly indicate whether the PF information provided is for the peak load time: for SPM (one of the two license areas of SPEN), we manually calculate the PF based on the provided active and reactive power during the peak load time; for SPD, the PF information is presented as a sub-column of the column "Maximum Load of Previous Year"; for ENW, the PF is presented as a column next to the column "Maximum Demand". We consider all these PF information as the PF over the peak load time. We find that 95% of PSs have PFs greater than 0.95, and the average PF over these PSs is 0.989 with a standard deviation of 0.0471. Given the high concentration of PFs, it is reasonable to assume an average PF equal to 0.989 for PSs without the PF information and for scaling the SSEN active power demand.

   For UKPN, the information provided is the demand headroom in percentage. Therefore, the peak demand in MVA net of the local generation can be calculated as:

   $$Net\ peak\ demand = Firm\ capacity\ (MVA) \times (1 - demand\ headroom\ \%)$$

   Note that UKPN data provides two sets of values for the firm capacity that reflect the impact of seasonality. We select the smaller one to reflect the stressful system condition.

   For NG, the peak demand in the required format can be calculated as:

   $$Net\ peak\ demand = Firm\ capacity\ (MVA) - demand\ headroom\ (MVA)$$

   It is noteworthy that SPM does not provide capacity information for individual PSs but for PS groups. Each of the PS group consists of one or interconnected several (up to 5) PSs. To address this, each row in our processed datasets corresponding to SPM is a PS group instead of a single PS. The peak net demand of the PS group is calculated by summing up the peak net demand of the individual PSs in that group.

The NPG data file [5] clearly mention the provided peak demand is at a half-hourly resolution. The time resolution of the peak demand data for other DNOs is mentioned in their methodology documents listed in Figure 2: The LTDS documents for ENW and SPEN, the "ssen-network-development-plan-methodology--assumptions.pdf" for SSEN, "NDP Methodology 2023 v1.0.pdf" for UKPN; "Network Capacity Report Methodology Document" for NG, suggest that the peak demand or the demand headroom are calculated based on half-hourly load curves.

2. LongLat for the locations of PSs. The information of ENW needs to be converted, using the conversion formula in EPSG or by using the `pyproj` package in `Python`. For SPEN, the interactive heat map [7] only provides the location of a single PS for each search based on the PS name. We manually checked all SPEN PSs to get the location information. For SPM, we set the location of each PS group based on the first PS within the group, considering the interconnected PSs within each group should be geographically close to each other.
3. Firm Capacity in MVA. Only the SSEN information needs to be processed, which can be calculated by the sum of the peak load (MVA) and the demand headroom (MVA). Note that, the SSEN "Availability Map" file provides the "Availability Headroom Capacity (MVA)", which is the only term without a clear definition in the data source. It is not clear whether this headroom is for demand or for generation. The other file, "SSEN Demand Headroom Report", provides the seasonal demand headroom forecasts for year 2023 made in 2022, which is clearly defined as the difference between the firm capacity and the net peak demand. The demand headroom recorded in this report (with an average of 7.95 MVA) is also much smaller than the "Availability Headroom Capacity (MVA)" (with an average of 10.80 MVA). Therefore, we instead look for the information recorded in the separate headroom report. We select the minimum headroom forecasts among all seasons to reflect the most stressful network condition.
4. Parent GSP. The GSP to which the PS is connected. GSP is the node that links the distribution grid to the transmission grid, providing valuable information to combine our processed distribution network datasets and other transmission-level network data in GB such as [8]. All the DNOs provide the required GSP information except SPM: although belonging to SPEN, source [9] only contains the GSP information for PSs of SPD but not for PSs of SPM; the GSP information of these SPM PSs is thus not included in our processed datasets.

| DNO | Data provided | Information-Contained | Date of download |
|---|---|---|---|
| ENW | Long-term Development Statement - load data [10] | PF<br>Peak demand (MVA)<br>Firm capacity (MVA) | March 2024 (Maximum demand for 2022/2023) |
|  | Network Headroom Report Data Workbook [11] | Location (X, Y)<br>Parent GSP | March 2024 |
| NPG | Northern Powergrid Capacity Heatmap [5] | Peak demand (MVA)<br>Firm capacity (MVA) | February 2023 |
|  | NPG NDP Demand Headroom [12] | Location (LongLat)<br>Parent GSP | July 2023 |
| SPEN (including SPD and SPM) | SPEN Long-term Development Statement (LTDS) [9] | PF<br>Peak demand (MVA)<br>Firm capacity (MVA)<br>Parent GSP (for SPD only) | February 2024 (Maximum demand for 2022/2023) |
|  | SP Distribution/Manweb Heat Maps [7] | Location (LongLat) | March 2023 |

| SSEN | Demand Headroom Report [13] | Demand headroom (MVA) | March 2023 (Demand forecasts of 2023 made in 2022) |
|---|---|---|---|
| | Availability Map [14] | Peak Demand (MW) Available Headroom Capacity (MVA) Location (LongLat) Parent GSP | March 2023 |
| UKPN | UK Power Networks PS area map [15] | Demand headroom (%) Location (LongLat) Firm capacity (MVA) | February 2023 |
| | Distributed Future Energy Scenarios Network Scenario Headroom Report [16] | Parent GSP | February 2024 |
| NG | Network Capacity Map [17] | Firm capacity (MVA) Demand headroom (MVA) Location (LongLat) Parent GSP | February 2023 |

*Table 1 Data sources of information on PSs.*

## Merging Datasets

From Table 1, we can see that some DNOs provide our required information in two separate files. As the next step, we merge the data of each DNO (if in separate files) and finally merge all DNOs' data to get the network dataset for the whole GB. The merging process is carried out by a **Python** script plus an additional manual check, which involves the following steps:

1. Pre-process PS names, including conversion to lower-case letters and the removal of spaces at the start and the end of the names.
2. A for-loop that enumerates each PS name of one data file. For each enumerated name, try to find the exact match in the other data file of the DNO. If the match is found, then we append the information; otherwise, we gradually remove the last character in the name and see if the match can happen. The intuition is that some PS names may contain meaningless suffix, such as "primary" (the type of the substation), or even some typos. However, this method may also make mistakes when the suffix is meaningful. Therefore, the script will print out the matched name of the PS for further manual validation. If dropping a certain number of characters still does not lead to a match, a manual scanning of the two datasets will be carried out again. If the match is still not found, we stop and leave the corresponding information empty.

Table 2 lists all the pairs of PS names that are matched after dropping characters and are manually validated. The pair is displayed in the format (PS Name1: PS Name 2), and different pairs are separated by a semicolon (;). NG only has a single data file; there is no need to have the matching process, so the corresponding entries in Table 2 are marked as not applicable (N/A). However, note that there are 70 out of 1,087 PSs of NG with missing information on firm capacity and peak demand, and these 70 PSs are dropped. For SPEN, there are only PSs of which the location information is not found in source [7], and there are no PS pairs that are matched through the iterative character-dropping process.

Among those matched and manually validated PS pairs in Table 2, we can see that some PS names are inconsistent simply because of abbreviations, such as "sq" and "square". Some PS names have suffix possibly for transformer indices, such as "t14" and "t13", and others have PS names grouped together to a single entry, such as "boroughbridge/roecliffe industrial". For ENW and NPG, we only append the information of locations and parent GSPs to the other file storing the demand and capacity, so the matches in the latter two cases are considered reasonable. For SSEN, there are inconsistencies among spellings (probably typos) among the matched pairs, such as "friockheim: friokheim" and "airigh-na-brodaig: arigh-na-brodaig", the extra slash (/), extra voltage information such as "11 kv", extra spaces between characters, or extra suffices "primary" at the end of the PS names. These inconsistencies are meaningless, so we consider these matched pairs reasonable as well. For SSEN and UKPN, there are other non-trivial inconsistencies, such as "lubnaig: loch lubnaig", "lairg grid: lairg", and "grovehurst local: grovehurst". In this case, we have manually verified that the matched PS name is the only overlapping name in both data files.

Finally, PSs with missing information on all attributes except the parent GSP and the PF are dropped. Among the 4,339 PSs recorded in GB DNOs, 4,205 of them are kept after the matching process and the final dropping. This leads to a minor loss of PS records (3.09%), but achieves the minimum amount of missing information in our datasets.

| DNO | Matched PS Names | Unmatched PSs |
|---|---|---|
| ENW | circle sq: circle square; | manchester university; samlesbury enterprise zone; st johns; thorley lane |
| NPG | boroughbridge/roecliffe industrial: boroughbridge; boroughbridge/roecliffe industrial: roecliffe industrial; sessay bridge_sowerby: sessay bridge; sessay bridge_sowerby: sowerby; fawdon t1 & seaton burn t1: fawdon t1; fawdon t1 & seaton burn t1: seaton burn t1; mansfield road 1: mansfield road | ottringham road; grimsby docks; yorkshire water hull road |
| SPEN | N/A | kingsland; brd t1; lostock t1 |
| SSEN | westpark fergus: west parkfergus; friockheim: friokheim; airigh-na-brodaig: arigh-na-brodaig, tarbert jura: tarbert, jura; tarbert loch fyne: tarbert, loch fyne; glenetive: glen etive; dervaig: dervaig (incl tiree); m v e e: mvee; winterbourne kingston: winterborne kingston; lubnaig : loch lubnaig; stoneywood t3: stoneywood mills t3; laburnum road: laburnam road; tarbert: tarbert, harris; bleddington: bledington; laburnam road: laburnum road; beaconsfield (end bb): beaconsfield (end b/b); beaconsfield (middle bb): beaconsfield (middle b/b); chandlersford: chandlers ford; chilton cantello: chilton cantelo; hinchelsea: hincheslea; crookham (church): crookham; mill lane, ringwood: mill lane; | plessey swindon; plessy titchfield; warminster; whiteley; chaorach; craigton; cromalt; Dounreay; grudie bridge; mill rd Montrose; phead north st; salen; salen 2; st fergus gas term; st fillans p s; stoneyhill; stoneywood t1&t2 |

| | south berstead: south bersted;<br>whiteley wood: whiteley;<br>yatton keynall: yatton keynell;<br>ardersier: arderseir;<br>burgarhill: burgar hill;<br>dunvegan grid: dunvegan;<br>ellon 11 kv: ellon;<br>glendoe primary: glendoe;<br>harbour: perth harbour;<br>inverness 11 kv: inverness;<br>kilmelford ps: kilmelford;<br>kishornhill: kishorn hill;<br>lairg grid: lairg;<br>quoich primary: quoich;<br>sandwick: setter sandwick;<br>st marys: st mary's;<br>dorchester town: dorchester;<br>pressed steel swindon se: pressed steel swindon;<br>bilbohall: bilbohal;<br>boat of garten 11 kv: garten;<br>bridge of dun 11kv: bridge of dun;<br>peterhead grange 11kv: peterhead grange | |
|---|---|---|
| UKPN | grovehurst local: grovehurst;<br>carnaby st 11 kv: carnaby st 11;<br>coulsdon 33/11kv: coulsdon;<br>morehall 132/11kv: morehall;<br>horley 33/11: horley | N/A |
| NG | N/A | N/A |

*Table 2 The manually matched PS name pairs in the format PS Name1: PS Name 2. For ENW and NPG, PS Name 1 is in the file storing the geo location and the parent GSP, while Name 2 is for the remaining information. For SSEN, PS Name 1 is in the file providing the demand headroom information only, and PS Name 2 is in the file providing the remaining information. For UKPN, PS Name 1 is in the file storing the parent GSP, while Name 2 is for the remaining information*

## Derivation of the Extended Dataset

The location information of each PS creates the possibility of linking to other social datasets, such as the census dataset. Here, we showcase the process of linking PSs to the number of households and the household heating composition data. One may follow a similar procedure to integrate other social datasets.

### Number of households

The number of households connected to each PS can be estimated assuming a proportional relationship between the number of households and the PS peak demand. In the real world, other factors also contribute to PS demand, such as 1) weather, which affects the electric demand for heating in winter, 2) local generation contributions, and 3) commercial and industrial connections. These factors will not contribute the same across the whole nation, but it is more reasonable to assume the same contribution within a relatively smaller geographical region. Therefore, we can first infer the corresponding sub-national region of GB for each PS, and then assign the number of households to the PS in proportion to the peak demand given the total number of households in that region. For example, if "PS a" (peak demand 8 MW) and "PS b" (peak demand 2 MW) are the only two PSs located in the sub-national region A with a total number of households being 10,000. Then it is estimated that "PS a" has 8,000 connected households while "PS b" has 2,000 household connections. Note that the area of the region should not be too small as well, as then a single PS can supply multiple regions, which means that our previous process assigning only one region to a PS is inaccurate.

Here the selected sub-national regions are 1) the 32 local authorities (LAs) for Scotland, the highest-level census geography that is exactly fitted [18], 2) the 9 government regions for England, which is the highest tier of subnational division in England in the Census data [19], and 3) the whole of Wales, considering its comparable area to a single government region in England. The corresponding region for each PS can be inferred based on its geographical location and the region boundary data recorded in [20].

### Residential Heating Composition

Given the estimated number of households connected to each PS, we find that the mean (median) number of connected households for each PS in England and Wales is 7,152 (respectively, 6,126). For Scotland PSs, the mean (median) number is 3,206 (respectively, 2,878). In census data, the geographical levels closest to the number of PS-level households are the Middle Layer Super Output Area (MSOA) level for England and Wales, and the Intermediate Zone (IZ) level for Scotland. The mean (median) number of households for an MSOA is 3,412 (respectively, 3,303). The mean (median) household number of an IZ is 1,855 (respectively, 1,802). Therefore, a PS can roughly support two MSOAs in England and Wales, or two IZs in Scotland. This observation suggests that the household heating composition of PS $i$ can be derived by combining the household heating data of the two nearest MSOAs in England and Wales, and of the two nearest IZs in Scotland.

The heating component information specifically refers to the number of households under each of the heating categories as listed in Section The Extended Dataset. Census data provide information at the IZ level in Scotland [18], and at the MSOA level in England and Wales [19]. Denote the connected number of households for PS $i$ as $N^{h,i}$. Suppose that we are estimating the number of households connected to the PS using the heating option $a$, denoted as $N^{a,i}$. Knowing the number of households with heating option $a$ in the two nearest MSOAs/IZs denoted as $N^{a,i,1}$ and $N^{a,i,2}$ respectively, we have:

$$N^{a,i} = N^{h,i} \times \frac{N^{a,i,1} + N^{a,i,2}}{N^{h,i,1} + N^{h,i,2}}$$

Where $N^{h,i,1}$ and $N^{h,i,2}$ stand for the total number of households in the two nearest MSOAs/IZs.

Finding the two nearest MSOAs/IZs for each PS can be efficiently carried out by the k-nearest neighbour algorithm based on the location of the PS and the location of the population centroids of MSOAs/IZs provided in [20] [21].

Finally, we need to ensure the consistency between the household heating information of PSs and that in the census data. In other words, the total estimated number of households in each heating option summed over PSs in GB (denoted as $N^{a,PS}$) should be equal to the total number of households in that heating option of GB loaded in the census data (denoted as $N^{a,Census}$). To achieve this equality, we multiply the estimated $N^{a,i}$ by $\frac{N^{a,Census}}{N^{a,PS}}$ for each PS. However, this can lead to slight mismatch between the total number of households connected to each PS summed over the heating options (estimated here) and that estimated in Section Number of households. In our case, the mean absolute percentage error between the two estimates is 1.52%. This may be an acceptable mismatch, but the data should be used carefully depending on the application. One could alternatively consider summing up the numbers of households across all the heating options to get the new estimated number of total household connections for each PS for better consistency. However, this makes the number of connected households of each PS not perfectly proportional to the PS peak demand, the assumption made to estimate the number of households in Section Number of households. The trade-off depends on whether the total number of households or the number of households for each specific heating option is of more interest. We have released the code in the file "link2social_data.ipynb", which facilitates modification in deriving the estimation in need.

# LIMITATIONS

The information on the parent GSPs of the 336 PSs of SPM is not found in our listed sources, which may create difficulties in connecting the PSs of SPM to other transmission network datasets. However, the remaining 3,869 PSs (92% of the total 4,205 PSs) in GB are linked to their parent GSPs in both of our released datasets (GB_PS_data.csv and GB_PS_data_extend.csv). The location and the corresponding geographic area of all the PSs are included in the file "GB_PS_data_extend.csv", based on which the 336 PSs of SPM could at least be approximately matched to nearby GSPs.

There are also some limitations in the extended dataset (GB_PS_data_extend.csv). First, the estimate of the number of connected households for each PS may be inaccurate, as the peak demand is affected by various factors, including industrial load and local generation connections, although we have tried to reduce this inaccuracy by only assuming the same effect of these factors in sub-national regions. In most cases, GB DNOs have information on the number of connected customers for each PS, but this information is not yet publicly available for most DNOs.

Second, as mentioned at the end of Section Residential Heating Composition, for each PS in our released extended dataset (GB_PS_data_extend.csv), there is a slight (1.52%) mismatch between the estimated total number of connected households and that derived by summing up the numbers of households in each heating option. We have discussed an alternative solution that can alleviate the mismatch but can also slightly violates the assumption made to estimate the number of households in Section Number of households. Codes are released to facilitate the modifications for its implementation.

# CRediT AUTHOR STATEMENT

**Yihong Zhou:** Conceptualization, Methodology, Investigation, Software, Data curation, Formal analysis, Writing, Original draft preparation, Validation. **Chaimaa Essayeh**: Data curation, Methodology, Writing-Reviewing and Editing, Validation. **Thomas Morstyn**: Conceptualization, Methodology, Writing-Reviewing and Editing, Supervision, Funding acquisition, Resources.

# ACKNOWLEDGEMENTS

The work was supported by the Engineering Studentship from the University of Edinburgh and was also supported by the UK Engineering and Physical Sciences Research Council (EPSRC) (Project references EP/S031901/1 and IAA PIV080).